\documentclass[reprint,amssymb,amsmath,,showpacs,preprintnumbers, prl]{revtex4-1}

\usepackage{graphicx}

\begin{document}

\title{To Grow is Not Enough: Impact of Noise on Cell Environmental Response and Fitness}
\author{Nash Rochman$^1$, Fangwei Si$^2$ and Sean X. Sun$^{2,3}$}%
\affiliation{$^1$Department of Chemical and Biomolecular Engineering, $^2$Department of Mechanical Engineering and Biomedical Engineering, Johns Hopkins University}

\begin{abstract}

Quantitative single cell measurements have shown that cell cycle duration (the time between cell divisions) for diverse cell types is a noisy variable. The underlying distribution is
mean scalable with a universal shape for many cell types in a variety of environments.
Here we show through both experiment and theory
that increasing the amount of noise in the regulation of the cell cycle negatively impacts the
growth rate but positively correlates with improved cellular response
to fluctuating environments. Our findings suggest that even non-cooperative
cells in exponential growth phase do not optimize fitness through
growth rate alone, but also optimize adaptability to changing conditions. In a manner similar to genetic evolution,
increasing the noise in biochemical processes correlates with improved response of the system to environmental changes.

\end{abstract}


\maketitle

The mantra,``Survival of the Fittest,'' coined by Spencer and popularized
by Darwin himself \cite{lyell1863geological}, pervades every corner
of biology. Fitness is usually defined to be the``birth-rate'' or the rate at which new individuals
are added to the population. Cooperative and mutlicellular systems may require a more complicated
definition; but often even these phenomena are shown to derive from the maximization of total sustainable
single-cell number \cite{libby2014geometry,segota2014spontaneous,hammerschmidt2014life,an2014bacterial}.
In the case of non-cooperative, single-cell species
(e.g. bacteria at low cell density), fitness as birth-rate is accepted.
For such a population during exponential growth, the number of cells
in an ensemble can be well described as a function of time if we know 
the initial number $N_{0}$, and the cell cycle duration $\tau$, 
yielding $N(t)=N_{0}\exp({\ln(2)} t/\tau)$.
In this way the constant $r={\ln(2)}/{\tau}$, often labeled the``growth-rate'', is used to measure fitness
- the larger $r$ and the faster an organism grows, the fitter it is.

However, the growth rate for a single cell is often hard to define. Experiments conducted in constant environments maintained in microfluidic devices (so called ``Mother Machines") show that the cell cycle duration \cite{wang2010robust} is stochastic and exhibits large variations for both prokaryotes and eukaryotes \cite{stukalin2013age}. Thus one should  consider a statistical distribution of cell cycle durations $P\left(\tau\right)$, where $\tau$ is the time between 2 successive cell divisions (septum formations). Owing to the fact that synthesis of new proteins and replication of DNA require finite time, there is a physical lower limit for the cell cycle duration, $\tau^*$ (dependent on the environment), below which no cells can divide. From an evolutionary perspective, we quickly see that to optimize fast growth, $P(\tau)$ should be a narrow distribution centered as close to $\tau^*$ as possible; however, the measured distribution for {\em E. coli} stands in stark opposition to this idea \cite{wang2010robust,stukalin2013age} (Fig. 1), exhibiting a significant variance in $\tau$. Quite strikingly, $P(\tau)$ is mean scaleable \cite{iyer2014universality,iyer2014scaling}  across a wide variety
of conditions, with shape conservation spanning cell types from {\em E. coli} to human dermal fibroblast cells \cite{stukalin2013age}. In Fig. 1 we display the distributions
and corresponding statistics for the ensembles investigated in this
work  and verify that they reflect the features discussed here. 

These observations, in conjunction with established cell cycle models
\cite{novak2008design,li2008quantitative} and more recent experimental results for protein synthesis and volume regulation,
have given rise to the present discussion about whether
a cell is best described as regulating its time until division, a``timer'' mechanism; volume at division, a``sizer'' mechanism;
or mass added over a single generation, a ``constant adder'' mechanism \cite{taheri2014cell,campos2014constant,amir2014cell}.

\begin{figure}
\begin{center}
\includegraphics[scale=0.45]{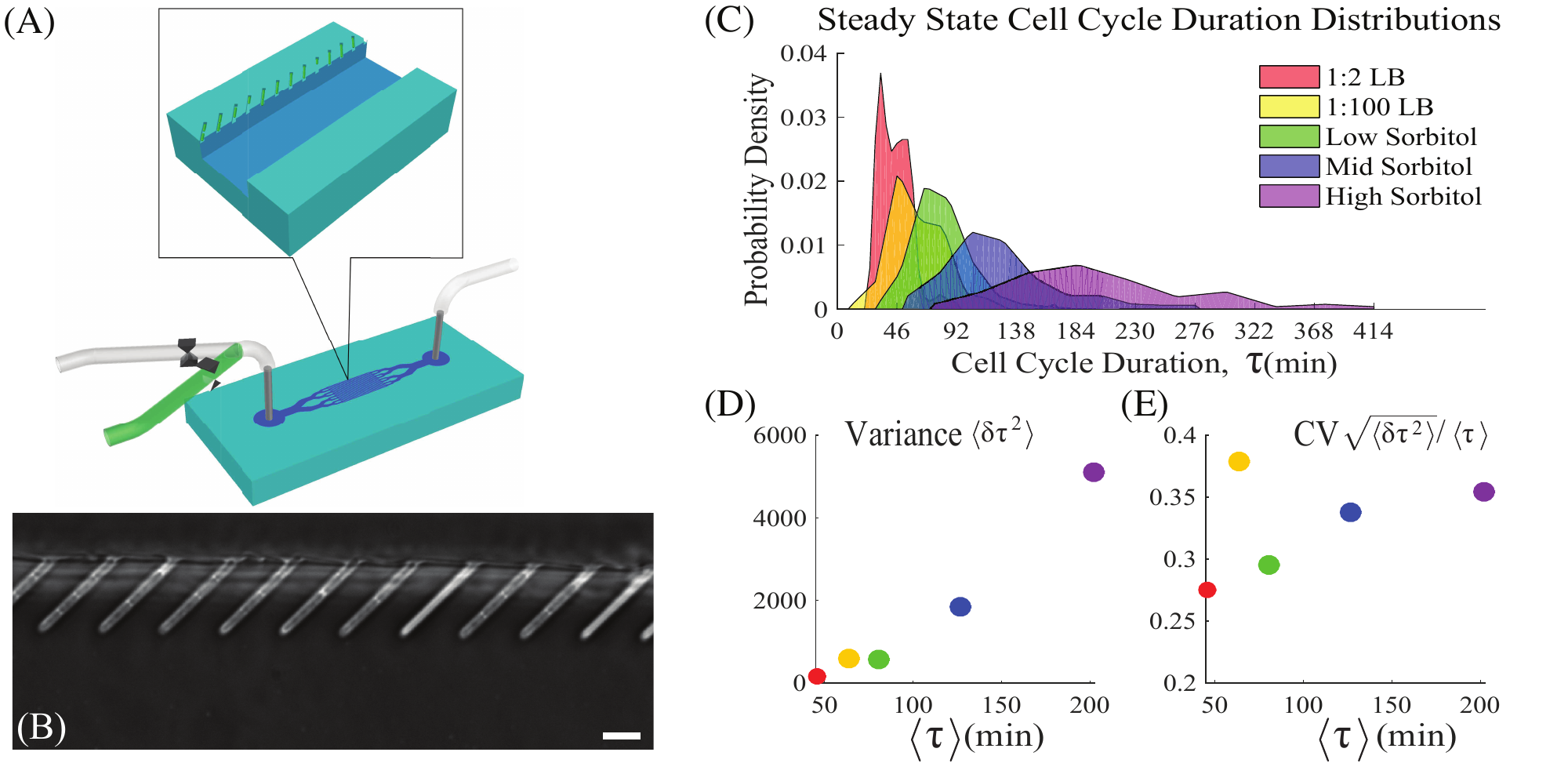}
\caption{\em (A) Cartoon of the Mother Machine (B) An image displaying {\em E. coli} cells in the microchannels. The scale bar is 5 microns. (C) {\em E. coli} cell cycle duration distributions (CCDDs) measured at constant nutrient conditions. See the Supplementary Materials (SM)
section III for more details. (D) Measured variation in the cell cycle duration and the Coefficient of Variation (CV): CV is roughly constant across all conditions.}
\end{center}
\end{figure}

Nevertheless, stochasticity in the cell cycle dynamics and heterogeneity in cell growth rate seems universal, which begs the question ``Why is this important?'' Specifically, we wish to probe the role
of programmed non-genetic heterogeneity apparent in this trend: that slow
growing cultures exhibit greater variability in their cell cycle regulation. The question
we will focus on for this investigation is, ``does increasing
the noise in the mechanisms regulating the cell cycle correlate with improved cell fitness?''

We first sought to determine how the variance of the cell cycle duration distribution (CCDD) correlates with the mean ensemble growth rate. The duration
distribution is best described as a shifted gamma distribution:
$\frac{1}{\Gamma(K)\theta^{K}}\left(\tau-\tau^{*}\right)^{K-1}e^{-\frac{\tau}{\theta}}$,
where $\theta$ and $K$ are parameters. The mean growth rate $r$ of the population is obtained by solving $\tau^* r+K ln\left(1+r/\theta\right)=\ln 2 $ \cite{stukalin2013age}.
$P(\tau)$ is bounded on the left due to the finite time required
to construct a new cell as discussed above. Given this minimum time
$\tau^*$, one may calculate the maximum growth rate for a given variance $\langle\delta\tau^2\rangle$ by changing parameters $(K, \theta)$.
In Fig. 2, we set $\tau^*$ to be length of the shortest observed
cell cycle and numerically calculated the maximum growth
rate for variances ranging over experimentally observed values.
Over this range, the maximum growth rate diminished by a factor of
three as we increased $\langle \delta \tau^2\rangle$. Clearly, increased noise in the regulation of the cell cycle correlates with
a lower growth rate. It is also important to note that cell division dynamics in {\em E. coli}
appears to be {\em ergodic}: each individual mother cell explores the entire distribution
and if data from a single cell is collected over a long period of
time, the resultant distribution appears to match that of a collection of
many cells at a single time, shown in Fig. 2. This suggests that there are no ``persistor cells'' that grow very slowly at all times to benefit the collective culture when subjected to harsh environments. 
More generally, it can be shown that given any CCDD
with finite width, there exists a narrower one which attains the same
growth-rate or greater (SM section I). We note that the existence of ``persistor cells"
has been confirmed \cite{deris2013innate,avery2005cell,balaban2004bacterial,lambert2015quantifying} in specialized cases and that these cells play an important
role in culture survivability. From our analysis of division phenotypes in the mother machine, however, they do not contribute significantly to the measured CCDD. 
\begin{figure}
\begin{center}
\includegraphics[scale=0.29]{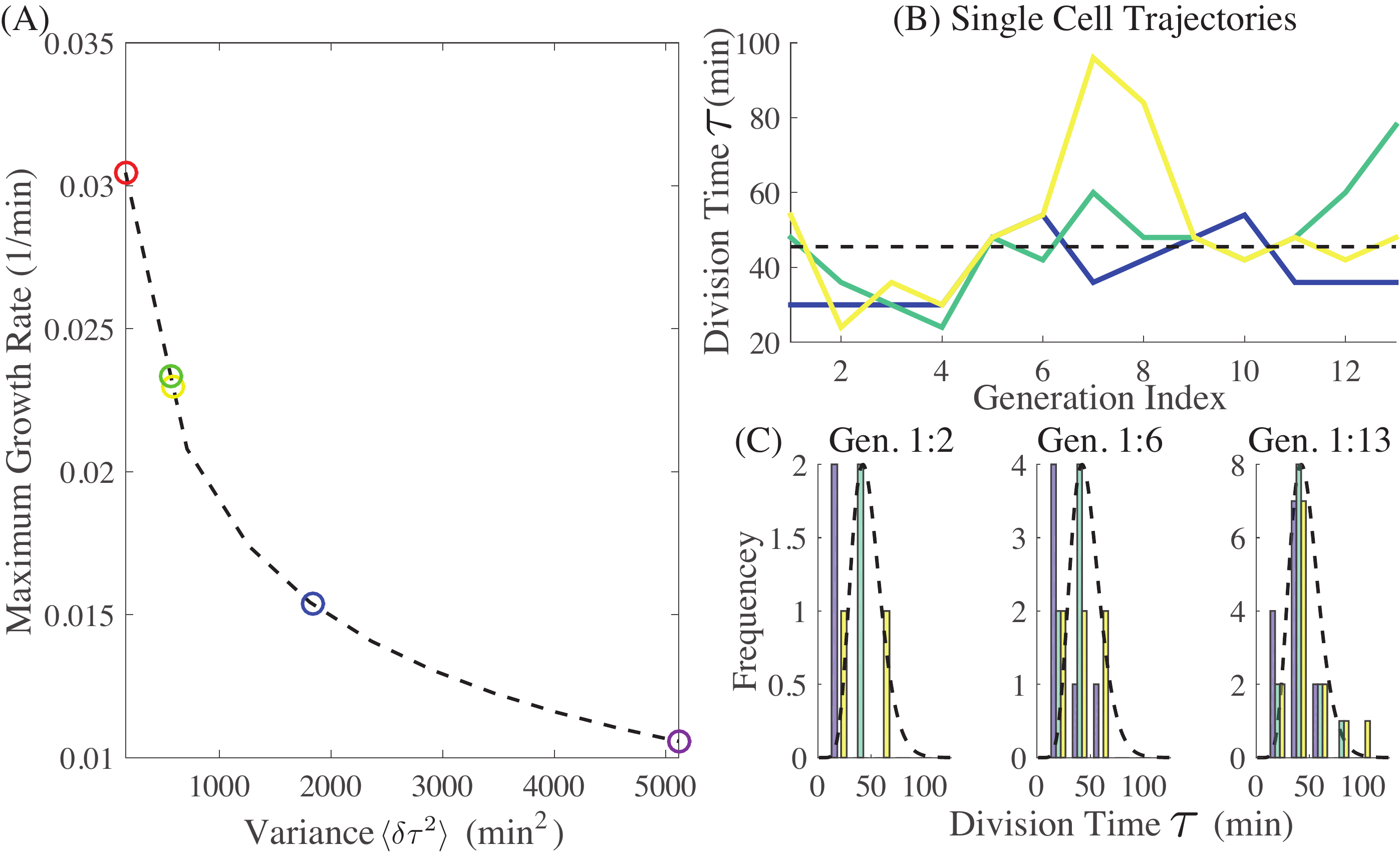}
\caption{\em (A) The maximum ensemble growth rate for $\tau^* = 12$min (the shortest
division recorded), as a function of variance (experimentally observed values are circles). (B) Single cell cycle duration trajectories (from 3 separate cells in the mother machine). The dotted line is the average duration. (C) The CCDD histogramed from single mother cells over different generation intervals (dotted line is the full distribution).}
\end{center}
\end{figure}

Thus, the population growth-rate is not improved by increasing the noise in the cell cycle
duration distribution and from the usual definition of fitness, this would suggest
cells should narrow this distribution. If this noise is intrinsic to underlying molecular mechanisms; however, and stochasticity is prevalent in gene expression \cite{elowitz2002stochastic,ray2012interplay}, polymerase activity \cite{raj2006stochastic}, and chemotaxis \cite{avery2006microbial,frankel2014adaptability}, cells might require higher energy consumption (sacrificing energy efficiency which carries its own evolutionary
importance \cite{Arijit2015bacterial}), or an increase of $\tau^*$ (increasing mean duration) to minimize noise. Here we suggest that there is another role (benefit) for the observed programmed non-genetic heterogeneity beyond the difficulty associated with noise reduction.

Organismal survival depends on two broad qualities - stability
and adaptability. Stability is a measure of short-term fitness, how
precisely a system can maintain conditions optimized for a constant
environment. Adaptability is a measure of long-term fitness, how quickly
a system is able to achieve optimized conditions when introduced to a
new environment. In the context of genetic evolution, a more mutable genome offers
an organism less stability but improved adaptability over many generations both through simply allowing for 
greater genetic diversity at any given time \cite{booy2000genetic,lacy1997importance} and high mutation rates
\cite{denamur2006evolution,moxon1994adaptive}
Here we demonstrate that the case is similar for the non-genetic noise between mother
and daughter cells on a much shorter timescale. 

To see how heterogeneity correlates with adaptation to environmental changes, 
we grew {\em E. coli} in the Mother Machine and collected single cell cycle duration data. We 
grew cells in five different types of media, performed step changes in the
growth medium, and measured how cells responded to these sudden environmental changes. 
Fig. 3 shows how the CCDDs evolved over time. For each experiment, the 
distribution is initially constant and stable before the sudden environmental change. After the change,
the distribution shifted over time, and eventually reached the new
stable distribution for the new environment. We find that the cell cycle duration
trajectories for individual cells follow a similar trend but include significant noise (Fig. 3 insets).

\begin{figure*}
\begin{center}
\includegraphics[scale=0.35]{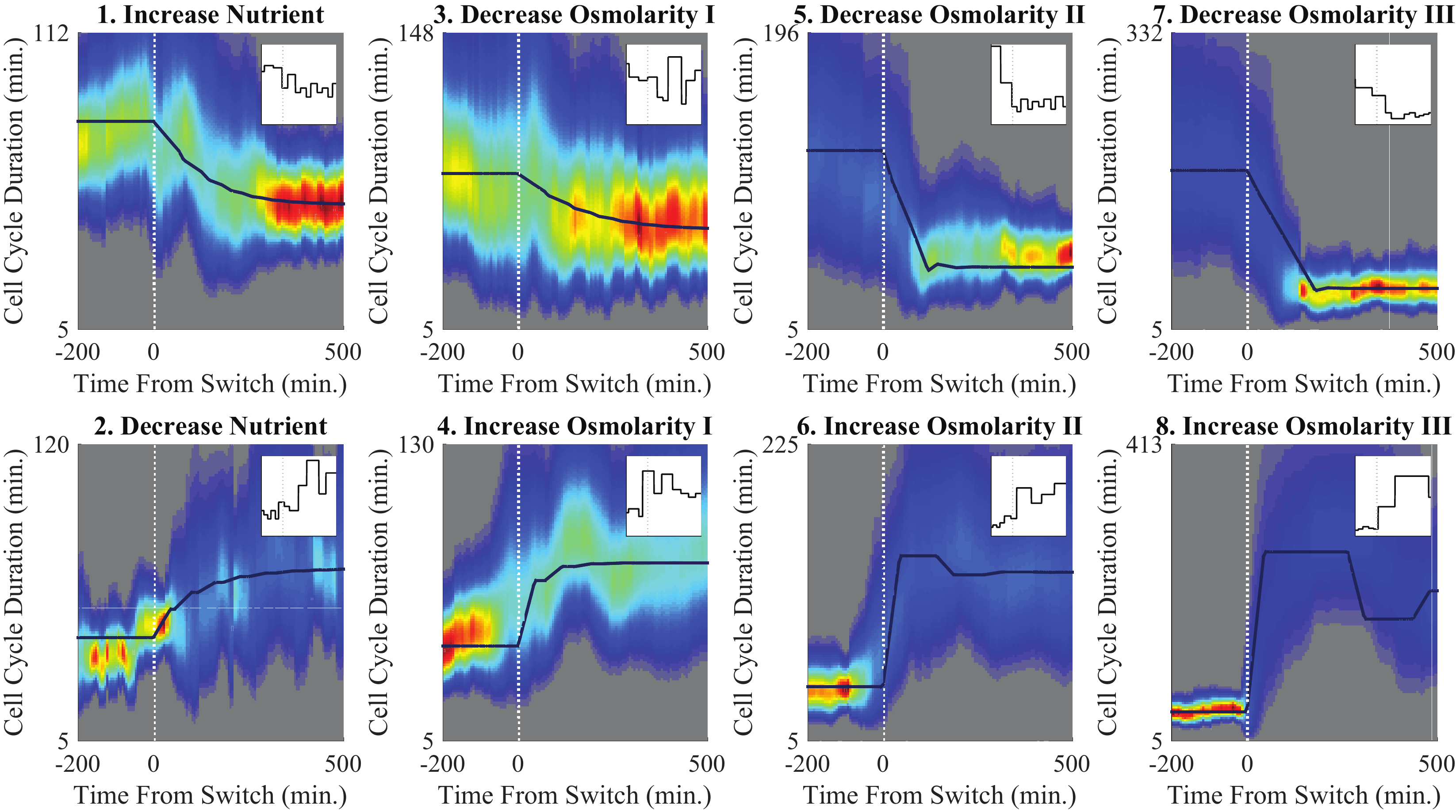}
\caption{\em Eight step environmental change experiments. The experimental distributions
(here fitted to Gamma distributions and smoothed; see Figure S5 for a direct comparison with the raw data) are displayed using colors with highest probability in red and lowest probability in blue. The black lines are the model predictions for the average. See SM section IV.}
\end{center}
\end{figure*}

We observed the response of CCDDs $\rho(\tau,n)$ in a series
of step change experiments where for time $t<0$ cells were
exposed to a constant environment, and at $t=0$ the environment was altered in such a way
that the new mean cell cycle duration was measurably different. (See SM section III).
The results are for {\em E. coli} in a  Mother Machine, but we also conducted a series of bulk
temperature shift experiments to compare with the microfluidic results
(See SM Section V). Fig. 3 shows the results of the eight environmental shift experiments
conducted. The top row contains relaxation experiments where the cells
were grown in suboptimal conditions including one nutrient limited and three hypertonic solutions
(where the osmolarity was increased with the addition of Sorbitol) before shifting the media
to the optimal environment (in diluted LB medium) for fast growth. The bottom row are the
reverse, stress experiments. The model is able to predict
the time course of cell response, including the overshoots observed in severe stress conditions (Fig. 3)

The trend across the top row in Fig. 3 is clear: as the magnitude of the shift increases (left to right), the response speed (change in the mean divided by the time over which the change occurred) increases
as well. A complementary trend may be observed on the bottom row:
as the severity of the stress increases (left to right), the response speed also increases. (See Fig. 4A).
However, while for the relaxation experiments (top row) response speed directly correlates with adaptability, 
for the stress experiments (bottom row) this is not the case. We consider an adaptable cell to be one which responds ``efficiently" to environmental changes in terms of its growth rate alone. (Here we observe negligible filamentation rates and cellular aging is not an issue as cells
were only followed for fewer than fifty generations.) In other words, its growth rate is as high as possible for as long as possible. In the case
of the relaxation experiments (top row), the faster the response speed, the greater the adaptability as a greater response speed allows the cell to spend more time in a fast growth state. For the stress experiments (bottom row) the reverse is true: the greater the response speed, the less time the cell is able to remain in the fast growing state. In fact, we have observed that for severe environmental stresses, cells respond so inefficiently that they attain a growth rate during the period of response which is even lower than that of the final stable growth rate (e.g. Fig. 3 last panel). Thus, to compare the efficiency of response across all experiments, we proposed to use a new quantity, $\Delta$, 
\begin{eqnarray}  \label{delta}
\Delta=\left \langle\frac{1}{\lambda} \int_0^\lambda\tau(t)dt \right\rangle -{\rm max}(\mu_i,\mu_f)
\end{eqnarray}
where the environment step change occurred at $t=0$, $\lambda=500$ min. is the minimum period for all eight experiments to complete the response to their new environments, and the average $\langle\rangle$ is taken over all cells. (Note that the trends observed are maintained over a wide range of $\lambda$. See SM section IV.) $\mu_{i,f}$ is the steady state average cell cycle duration before and after the step change, respectively.  When $\Delta$ is large and positive, the cells respond so inefficiently that during response they grow even slower than in the stressed condition. When $\Delta$ is large and negative, the cells are able to remain or enter in the fast growth state for the majority of the response period. We analyze these results in the context of a phenomenological model introduced below.

We consider a sequence of cell cycles $(\tau_1,\tau_2,\dots)$ and the evolution of the cell cycle distribution over generations, $\rho(\tau, n)$ where $n$ is the index of generation. For a constant environment at long times, cells are at steady state in the Mother Machine, and $\rho(\tau,n)\equiv P(\tau)$ shown in Fig. 1. For a changing environment, we consider a
Markovian stochastic model for $\rho(\tau,n)$, which describes the evolution of this distribution from one generation to the next as
\begin{eqnarray}
\rho(\tau_i,n+1)= \int d\tau_j M(\tau_j \rightarrow \tau_i; \phi) \rho(\tau_j,n)  \label{dynamics}
\end{eqnarray}
where $M$ is the transition probability, which depends on the current environment described by $\phi$. $M$ describes the probability of a daughter cell to divide after duration $\tau_i$ given that the mother cell divided after duration $\tau_j$. Eq. \ref{dynamics} is simply a statement of probability conservation; and by developing a model for $M$, we can predict how cells can respond to environmental changes over time.


Motivated by these ideas in conjunction with the ``constant-adder
model'' and older foundational work \cite{bremer1981cell}, we propose an approximate Gaussian model for the 
cell cycle dynamics:
\begin{eqnarray}
M(\tau_{j}\rightarrow\tau_{i}) &\propto& \exp \left[-\frac{(\tau_{i}+ \alpha \tau_{j}-(1+ \alpha)\mu (\phi) )^{2}}{2\sigma_{1} (\phi) ^{2}}\right] \nonumber \\
&& \times \exp \left[-\frac{(\tau_{i}-\tau_{j})^{2}}{2\sigma_{2} (\phi)^{2}}\right]
\end{eqnarray}
where $\sigma_{1},\sigma_{2}$ are
positive constants describing the noise,  and $\mu$ is the mean cell cycle duration for a given condition $\phi$. $\sigma_1,\sigma_2$ and $\mu$ all depend on the current environmental variable $\phi$.  $\alpha$ is a constant between 0 and 1, representing the fraction of the cell cycle after DNA replication
has terminated and before septum formation. See SM
section II. The first term represents adaptability - increasing its weight (decreasing
$\sigma_{1}$) is equivalent to boosting the amount of information
a cell may obtain about its environment. Conversely, the second term
represents stability and increasing its weight (decreasing $\sigma_{2}$)
is equivalent to increasing the similarity between mother and daughter
cells. When the environment is constant, dynamics generated by $M$ must produce the steady state
CCDD, $P(\tau)$. Therefore, $(\sigma_1,\sigma_2,\mu)$ are determined by the environment and their corresponding steady state distributions. Indeed, for constant environmental conditions, the model predicts that the correlation of cell cycle duration is, 
\begin{eqnarray}
C(n)=\langle\delta\tau(0)\delta\tau(n)\rangle/\langle\delta\tau^2(0) \rangle = \left[\frac{1/\sigma_2^2 -\alpha/\sigma_1^2}{1/\sigma_2^2+1/\sigma_1^2}    \right]^n
\end{eqnarray}
Since $(\sigma_1,\sigma_2)$ depend on environmental conditions, this result is a way to use steady state cell cycle correlations to obtain $M$. In addition, it is possible to explicitly obtain transition probabilities from the experimental data. The comparison between our model $M$ and the data collected is shown in Fig. S1.   (see SM section IV)

Unlike the response speed which increases with increasing environmental shock severity over all experiments, we find that $\Delta$ for the shock and relaxation experiments display two opposing trends: as the severity of the environmental change, measured as $|\mu_i-\mu_f |$ increases, $\Delta$ decreases for the relaxation experiments and increases for the stress experiments (See Fig. 4B). Clearly the response efficiency cannot be predicted from the severity of the environmental change alone. According to our model, there is a parameter that should be well correlated with response efficiency, $\sigma_{2}$. As $\sigma_{2}$ (calculated from the initial state) increases, the similarity between mother and daughter cells decreases - which should make the cell more adaptable. We find that this agrees with experiment:
as $\sigma_{2}$ increases, $\Delta$ decreases (Fig. 4C). $\sigma_{2}$ is not the ideal parameter for comparison, however, since it cannot be directly measured experimentally. It would be better if the same trend could be observed
for the total variance, $\langle \delta\tau^2\rangle$, of the cycle duration distribution. Here we may utilize a result from the constant-adder model, which predicts that the autocorrelation function is conserved across different environmental conditions \cite{taheri2014cell}. Given the autocorrelation function we can derive $\sigma_{2}$ with the variance of the ensemble. As the variance increases, $\sigma_{2}$ increases (Fig. 4D). Thus knowing that $\sigma_{2}$ must be large for an efficient
response, then the variance must also be large for an efficient response. We find that this well agrees with experiment:
as the variance of the initial CCDD increases, $\Delta$ decreases (Fig. 4E).

\begin{figure}
\begin{center}
\includegraphics[scale=0.3]{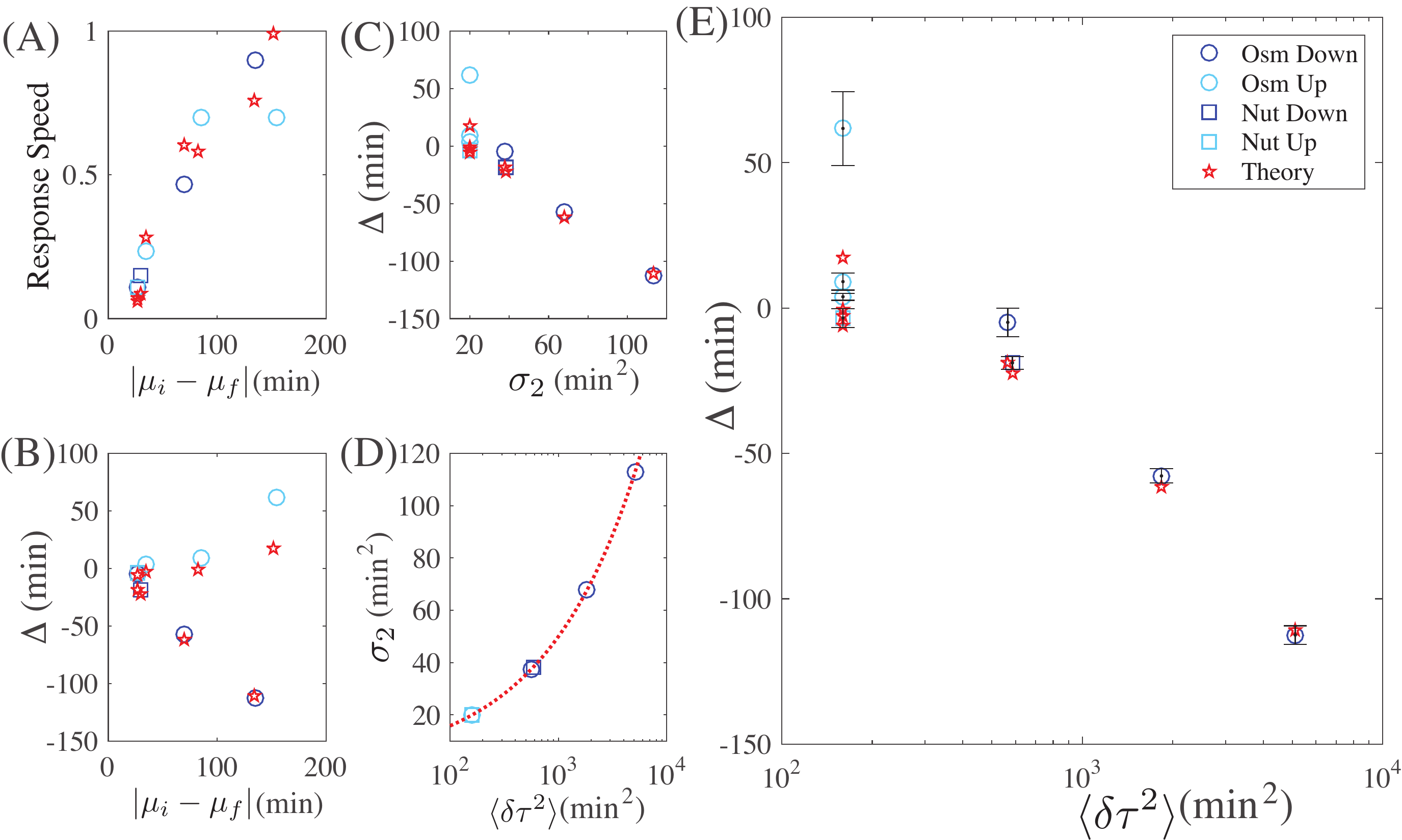}
\caption{\em (A) Average cell response speed vs shock severity: $|\mu_f-\mu_i|$. (B) $\Delta$ vs. shock severity $|\mu_f-\mu_i|$. (C) $\Delta$ vs $\sigma_{2}$ for the initial ensemble before the step change. (D) $\sigma_{2}$ vs. variance of the cell cycle duration $\langle \delta \tau^2\rangle$ derived from the autocorrelation function before the step change. (E) $\Delta$ vs variance (bars denote standard error). As the variance of the initial CCDD before the environment change increases, $\Delta$ decreases, and thus the response efficiency improves. Theory results are displayed as red lines or stars.}
\end{center}
\end{figure}

We may now return to answer our original question,``does increasing
the noise in the mechanisms regulating the cell cycle correlate with improved cell fitness?'' The answer appears to be yes: Increasing
the variance of the CCDD, attributable to increasing non-genetic heterogeneity of the culture,
correlates with improved adaptability of the cell to environmental changes. This is
because with increased noise, cells can explore a
wider range of phenotypes and some are already well suited for a new environment before it is introduced.
In other stochastic systems, the fluctuation dissipation  theorem (FDT) expresses a similar concept. However, we have not proved this connection conclusively, since in {\em E. coli}, $\langle \delta\tau^2 \rangle$ is also inversely correlated with the mean division time. One could also argue that a slower dividing cell responds efficiently. The conclusive proof requires comparison between strains that divide with the same mean, but different $\langle \delta\tau^2 \rangle$. Such a construct is currently not available to us.


On much shorter time scales, noise in protein expression \cite{swain2002intrinsic} has
proven to be important for cellular robustness \cite{selvarajoo2012non} and displayed clinical relevance:
increasing the stochasticity of protein expression can help combat dormant pathogens such as HIV \cite{dar2014screening}. Here we showed that these short term fluctuations in the biochemical regulation of the cell cycle are correlated with the ability of a cell to adapt to a changing environment, analogous to the long term genetic adaptations\cite{gonzalez2015stress} and complementary to long term memory of a periodic, flucuting environment\cite{kussell2005phenotypic,lambert2014memory}. Our results here provide a reason for the mean-scaling (or fixed CV) of CCDDs and their universal shape: when growing fast, cells benefit
most from stability where a greater gain may be achieved from optimizing
growth for the current and immediate environment; however when growing
slowly, cells benefit more from improving their adaptability so that
when superior growth conditions are presented they may respond efficiently
to best utilize the new environmental conditions. When the mean cell cycle duration is large,
it most benefits the population to be heterogeneous. When the mean is small it is best for
the population to be homogenous.


\begin{thebibliography}{99}

\bibitem{lyell1863geological}
Lyell, C.
\newblock (1863) {\em The geological evidences of the antiquity of man: with
  remarks on theories of the origin of species by variation}.
\newblock (J. Murray).

\bibitem{libby2014geometry}
Libby, E, Ratcliff, W, Travisano, M, Kerr, B,  \& Cordero, O.~X.
\newblock (2014) 
\newblock {\em PLoS Comput Biol} {\bf 10}, e1003803.

\bibitem{segota2014spontaneous}
Segota, I, Boulet, L, Franck, D,  \& Franck, C.
\newblock (2014) 
\newblock {\em Physical biology} {\bf 11}, 036001.

\bibitem{hammerschmidt2014life}
Hammerschmidt, K, Rose, C.~J, Kerr, B,  \& Rainey, P.~B.
\newblock (2014) 
\newblock {\em Nature} {\bf 515}, 75--79.

\bibitem{an2014bacterial}
An, J.~H, Goo, E, Kim, H, Seo, Y.-S,  \& Hwang, I.
\newblock (2014) 
\newblock {\em Proceedings of the National Academy of Sciences} {\bf 111},
  14912--14917.

\bibitem{wang2010robust}
Wang, P, Robert, L, Pelletier, J, Dang, W.~L, Taddei, F, Wright, A,  \& Jun, S.
\newblock (2010) \newblock {\em Current biology} {\bf 20}, 1099--1103.

\bibitem{stukalin2013age}
Stukalin, E.~B, Aifuwa, I, Kim, J.~S, Wirtz, D,  \& Sun, S.~X.
\newblock (2013) 
\newblock {\em Journal of The Royal Society Interface} {\bf 10}, 20130325.

\bibitem{iyer2014universality}
Iyer-Biswas, S, Crooks, G.~E, Scherer, N.~F,  \& Dinner, A.~R.
\newblock (2014) 
\newblock {\em Physical review letters} {\bf 113}, 028101.

\bibitem{iyer2014scaling}
Iyer-Biswas, S, Wright, C, Henry, J, Lo, K, Burov, S, Lin, Y, Crooks, G, Crosson, S, Dinner, A, \& and Scherer, N.
\newblock{2014}
\newblock {\em Proceedings of the National Academy of Sciences}{\bf 111(45)}, 15912--15917

\bibitem{novak2008design}
Nov{\'a}k, B \& Tyson, J.~J.
\newblock (2008) 
\newblock {\em Nature reviews Molecular cell biology} {\bf 9}, 981--991.

\bibitem{li2008quantitative}
Li, S, Brazhnik, P, Sobral, B,  \& Tyson, J.~J.
\newblock (2008) 
\newblock {\em PLoS computational biology} {\bf 4}, e9.

\bibitem{taheri2014cell}
Taheri-Araghi, S, Bradde, S, Sauls, J.~T, Hill, N.~S, Levin, P.~A, Paulsson, J,
  Vergassola, M,  \& Jun, S.
\newblock (2014) 
\newblock {\em Current Biology}.

\bibitem{campos2014constant}
Campos, M, Surovtsev, I.~V, Kato, S, Paintdakhi, A, Beltran, B, Ebmeier, S.~E,
  \& Jacobs-Wagner, C.
\newblock (2014) 
\newblock {\em Cell} {\bf 159}, 1433--1446.

\bibitem{amir2014cell}
Amir, A.
\newblock{2014}
\newblock{\em Physical Review Letters}
\newblock{\bf 112}, 208102

\bibitem{deris2013innate}
Deris, J.~B, Kim, M, Zhang, Z, Okano, H, Hermsen, R, Groisman, A,  \& Hwa, T.
\newblock (2013) 
\newblock {\em Science} {\bf 342}, 1237435.

\bibitem{avery2005cell}
Avery, S.~V.
\newblock (2005) 
\newblock {\em Trends in microbiology} {\bf 13}, 459--462.

\bibitem{balaban2004bacterial}
Balaban, N, Merrin, J, Chait, R, Kowalik, L \& Leibler, S.
\newblock{2004}
\newblock{\em Science} {bf\ 305(5690)}, 1622--1625

\bibitem{lambert2015quantifying}
Lambert, G \& Kussell, E.
\newblock{2015}
\newblock {\em Physical Review X} {\bf 5(1)}, 011016

\bibitem{elowitz2002stochastic}
Elowitz, M.~B, Levine, A.~J, Siggia, E.~D,  \& Swain, P.~S.
\newblock (2002) 
\newblock {\em Science} {\bf 297}, 1183--1186.

\bibitem{ray2012interplay}
Ray, J. C.~J. \& Igoshin, O.~A.
\newblock{2012}
\newblock{\em PLoS Computational Biology}
\newblock{\bf 8}

\bibitem{raj2006stochastic}
Raj, A, Peskin, C.~S, Tranchina, D, Vargas, D.~Y,  \& Tyagi, S.
\newblock (2006) 
\newblock {\em PLoS Biol} {\bf 4}, e309.

\bibitem{avery2006microbial}
Avery, S.~V.
\newblock (2006) \newblock {\em Nature Reviews Microbiology} {\bf 4}, 577--587.

\bibitem{frankel2014adaptability}
Frankel, N.~W, Pontius, W, Dufour, Y.~S, Long, J, Hernandez-Nunez, L,  \&
  Emonet, T.
\newblock (2014) 
\newblock {\em eLife} {\bf 3}, e03526.

\bibitem{Arijit2015bacterial}
Arijit, M. \& Dill, K.~A.
\newblock (2015) 
\newblock {\em Proceedings of the National Academy of Sciences} {\bf 112.2}, 12795--12800.

\bibitem{booy2000genetic}
Booy, G, Hendriks, R, Smulders, M, Groenendael, J.~v,  \& Vosman, B.
\newblock (2000) 
\newblock {\em Plant biology} {\bf 2}, 379--395.

\bibitem{lacy1997importance}
Lacy, R.~C.
\newblock (1997)\newblock {\em Journal of Mammalogy} {\bf 78}, 320--335.

\bibitem{denamur2006evolution}
Denamur, E \& Matic, I.
\newblock (2006) 
\newblock {\em Molecular microbiology} {\bf 60}, 820--827.

\bibitem{moxon1994adaptive}
Moxon, E.~R, Rainey, P.~B, Nowak, M.~A,  \& Lenski, R.~E.
\newblock (1994) \newblock {\em Current biology} {\bf 4}, 24--33.

\bibitem{bremer1981cell}
Bremer, H \& Chuang, L.
\newblock (1981) 
\newblock {\em Journal of theoretical biology} {\bf 88}, 47--81.

\bibitem{swain2002intrinsic}
Swain, P.~S, Elowitz, M.~B,  \& Siggia, E.~D.
\newblock (2002) \newblock {\em Proceedings of the National Academy of Sciences} {\bf 99},
  12795--12800.

\bibitem{selvarajoo2012non}
Selvarajoo, K.
\newblock (2012) \newblock {\em Frontiers in genetics} {\bf 4}, 287--287.

\bibitem{dar2014screening}
Dar, R.~D, Hosmane, N.~N, Arkin, M.~R, Siliciano, R.~F,  \& Weinberger, L.~S.
\newblock (2014) \newblock {\em Science} {\bf 344}, 1392--1396.

\bibitem{gonzalez2015stress}
Gonz{\'a}lez, C. \emph{et~al.}
\newblock{2015}
\newblock {\em Molecular systems biology}{\bf 11},827

\bibitem{kussell2005phenotypic}
Kusell, E \& Leibler, S.
\newblock{2005}
\newblock{\em Science} {\bf 309},2075--2078.

\bibitem{lambert2014memory}
Lambert, G \& Kussell, E.
\newblock{2014}
\newblock {\em PLoS Genetics}{\bf 10(9)}

\end{thebibliography}
\end{document}